\documentclass[%reprint,
 amsmath,amssymb,
onecolumn]{revtex4-2}

\usepackage{graphicx}% Include figure files
\usepackage{dcolumn}% Align table columns on decimal point
\usepackage{bm}% bold math
\newcommand{\sig}[1]{\bm{\sigma}_{#1}}

\DeclareMathOperator{\Tr}{Tr}

\begin{document}

%\preprint{APS/123-QED}

\title{Non-Hermitian Generalization of Bloch Sphere in Spacetime Algebra}% Force line breaks with \\

\author{Chih-Wei Wang}
 \email{freeform1111@gmail.com}
 \altaffiliation{Independent researcher}%Lines break automatically or can be forced with \\

\date{\today}% It is always \today, today,
             %  but any date may be explicitly specified

\begin{abstract}

We establish a geometric generalization of the Bloch sphere for two-level quantum systems with non-Hermitian Hamiltonians using the Spacetime Algebra (STA) formulation. By lifting the state density operator from the even subalgebra to the full STA, we show that the state space expands from the unit 2-sphere to a future light cone. The non-unitary time evolution generated by a general non-Hermitian Hamiltonian corresponds to proper orthochronous Lorentz transformations on the null vectors. We classify the Hamiltonian dynamics into four distinct geometric classes—spatial rotations (corresponding to $\mathcal{PT}$-symmetric systems), pure boosts (anti-$\mathcal{PT}$ symmetric systems), null rotations (exceptional points), and general mixtures. We also use this formulation to study several results from $\mathcal{PT}$-symmetric quantum mechanics including the topological features of the exceptional points.

\end{abstract}

%\keywords{Suggested keywords}%Use showkeys class option if keyword
                              %display desired
\maketitle

%\tableofcontents

%%%%%%%%%%%%%%%%%%%%%%%%%%%%%%%%%%%%%%%%%%%%%%%%%%%%%%%%%%%%%%%%%%%%%%%%%%%%%%%%%%%%%%%%%%%%%%%%%%%%%%%%%%%%%%%%%%%%%%%

\section{Introduction}

%The standard quantum mechanics formulation is based on a complex Hilbert space and matrix operations. However, several more intuitive and geometrical perspectives for some simple systems or more general formulations still exist. For example, it is well known that a pure state of a two-level quantum system (i.e., a spin-1/2 particle or a qubit) can be represented uniquely by a unit vector on the Bloch sphere. If the Hamiltonian is Hermitian, the time evolution of the state can be understood as a spatial rotation on the Bloch sphere. Furthermore, there are several attempts to generalize this picture to multi-qubit systems. 

%The standard formulation of quantum mechanics is built upon complex Hilbert spaces and matrix operations. However, intuitive geometric perspectives exist for lower-dimensional quantum systems. For example, a pure state of a two-level system (a qubit or spin-1/2 particle) can be uniquely represented by a unit vector on the Bloch sphere. When the Hamiltonian is Hermitian, state time evolution reduces to spatial rotations on this sphere. Various efforts have also been made to generalize this geometric picture to multi-qubit systems \cite{Wie:2020awo,pinero2026extending,shoji2026visualization}.

The standard formulation of quantum mechanics relies on complex Hilbert spaces and matrix operators. However, intuitive geometric formulations exist for lower-dimensional quantum systems. For instance, a pure state of a two-level quantum system (a qubit or spin-1/2 particle) is uniquely represented by a unit vector on the Bloch sphere. Under a Hermitian Hamiltonian, unitary time evolution corresponds to simple spatial rotations on this sphere. Various efforts have extended this geometric interpretation to multi-qubit systems through tensor products and higher-dimensional spheres \cite{Wie:2020awo,pinero2026extending,shoji2026visualization}.

An alternative geometric foundation developed by Hestenes utilizes geometric (Clifford) algebra \cite{gurtler1975consistency,havel2000geometric,doran2003geometric}. In this framework, the imaginary unit $i$ and quantum spinors receive concrete spatial interpretations. Hestenes further formulated Dirac theory \cite{hestenes1967real,hestenes1973local,hestenes2003spacetime,hestenes2003mysteries} using Spacetime Algebra (STA)—the Clifford algebra $C\ell_{1,3}(\mathbb{R})$ of Minkowski spacetime. Within STA, the density operator of a qubit takes a concise, coordinate-free form that mirrors the Bloch sphere. This formulation extends naturally to multi-qubit systems \cite{doran2003geometric,wang2018density} through Multiparticle Spacetime Algebra (MSTA) \cite{doran2003geometric}.

In standard quantum theory, Hermiticity is routinely imposed to ensure real energy spectra and unitary probability conservation.  However, it has been discovered that a Hamiltonian that is invariant under combined parity $P$ and time-reversal $T$ symmetry can possess entirely real spectra despite being non-Hermitian \cite{bender1998real}. Therefore, Hermiticity is a sufficient but not necessary condition to ensure real energy eigenvalues. This realization led to rapid developments in non-Hermitian and $\mathcal{PT}$-symmetric quantum mechanics. There is also a compelling theoretical and experimental argument that we should relax Hermiticity to a weaker condition: $\mathcal{PT}$-symmetry \cite{bender2002complex,mostafazadeh2003exact,bender2024pt}. However, if Hermiticity is relaxed, the Bloch sphere needs to be modified to incorporate the $\mathcal{PT}$-symmetric Hamiltonian or a more general non-Hermitian one. The main purpose of this paper is to find a way to generalize the Bloch sphere to incorporate the non-Hermitian Hamiltonian using the STA formulation.

%We started from a density operator of a qubit expressed in the even subalgebra of the STA. Then, we study the time evolution of the density operator under a Hermitian or non-Hermitian Hamiltonian. We found that to extend the Bloch sphere to the non-Hermitian case, we need to lift the even subalgebra to the full STA. The conclusion is that we need to replace the unit sphere with a future light cone, and the time evolution of the states corresponds to different types of Lorentz transformations depending on the Hamiltonian. 

We start from the density operator of a qubit defined within the even subalgebra of STA. Studying its time evolution under Hermitian and non-Hermitian Hamiltonians reveals that incorporating non-Hermitian dynamics requires lifting the representation to the full STA. As a result, the unit Bloch sphere expands into a future light cone, where time evolution corresponds to different types of Lorentz transformations depending on the Hamiltonian. 

%This paper is structured as follows: in section \ref{sec:STA}, we briefly review the essential part of STA, then in section \ref{sec:Bloch_non_Hermitian}, we use the STA formulation to study the non-Hermitian generalization of a Bloch sphere. In section \ref{sec:Types_Lorentz}, we categorize the Hamiltonian into four different kinds, and each kind corresponds to a certain type of Lorentz transformation. Finally, in section \ref{sec:discussion}, we talk about several unsolved questions and mention several future directions.

This paper is organized as follows: Section \ref{sec:STA} provides a brief review of STA fundamentals. Section \ref{sec:Bloch_non_Hermitian} develops the STA formulation for the non-Hermitian generalization of the Bloch sphere. In Section \ref{sec:Types_Lorentz}, we categorize the Hamiltonian into four different kinds, and each kind corresponds to a certain type of Lorentz transformation. Finally, in Section \ref{sec:discussion}, we discuss several open questions and point to several future directions.

\section{\label{sec:STA}Spacetime Algebra}

Spacetime algebra (STA) is a specific formulation of geometric algebra (or Clifford algebra) applied to the four-dimensional Minkowski spacetime introduced by Hestenes. It provides an elegant, coordinate-free approach to various areas of physics. In this section, we will only briefly review the necessary parts of the algebra for this paper; the details can be found in \cite{hestenes2003spacetime,doran2003geometric,hestenes2015space}.

\subsection{Basic properties of STA}

Spacetime algebra (STA) is generated by four orthonormal vectors: $\{\gamma_0\,,\gamma_1\,,\gamma_2\,,\gamma_3\}$ equipped with an associative operation between vectors called the geometric product. This product between two vectors $u$ and $v$ is denoted as $u\,v$ and can be decomposed into a symmetric part and an antisymmetric part:
\begin{equation}
    u\,v = u\cdot v ~+~ u\wedge v\,.
\end{equation}
The symmetric part is the usual inner product and yields a scalar:
\begin{equation}
    u\cdot v = \frac{1}{2}(u v + v u) = v\cdot u
\end{equation}
The antisymmetric part is the outer product and yields a bivector:
\begin{equation}
    u\wedge v = \frac{1}{2}(u v - v u) = - v\wedge u
\end{equation}

Adopting mostly minus signature of the spacetime, the basis vectors, $\gamma_\mu$, satisfy:
\begin{equation}
    \gamma_\mu\cdot\gamma_\nu = \eta_{\mu\nu}=\text{diag}(1,-1,-1,-1)\,.
\end{equation}
The unit pseudoscalar can be defined:
\begin{equation}
    I = \gamma_0\gamma_1 \gamma_2 \gamma_3\,,
\end{equation}
and it satisfies:$I^2=-1$ and anti-commutes with all spacetime vectors (i.e., $I \gamma_\mu = - \gamma_\mu I$). Furthermore, it can be shown that the pseudoscalar plays a similar role as a unit imaginary number, $i$, in standard quantum mechanics. 

The STA consists of 16 linearly independent elements:
\begin{equation}
    1\,, \quad \gamma_\mu,  \quad \gamma_\mu\wedge\gamma_\nu\,,  \quad
    I \gamma_\mu\,, \quad I\,.
\end{equation}
They are separated into five groups from grade-0 to grade-4 and can be referred to as $k$-vectors with $k=0,1,2,3,4$. Any element $M$ in the STA can be decomposed into its $k$-vector parts:
\begin{equation}
    M = \alpha ~+~ a ~+~ B ~+~ I b ~+~ I\beta\,,
\end{equation}
where $\alpha$ and $\beta$ are scalars, $a$ and $b$ are vectors and $B$ is a bivector. 

An important operation for the computation is the reverse operation denoted by the tilde $\tilde{}$:
\begin{equation}
    \tilde{I} = \gamma_3\gamma_2\gamma_1\gamma_0 = I\,.
\end{equation}
This operation reverses the order of the geometric products. Furthermore, the effect of the reverse depends on the grades of the terms. For example, the reverse of $M$ is:
\begin{equation}
    \tilde{M} = \alpha ~+~ a ~-~ B ~-~ I b ~+~ I\beta\,,
\end{equation}

\subsection{Spacetime splits and relative space}

While the STA allows formulas to be written in a fully covariant, coordinate-independent manner, the results are usually expressed in a specific frame. Therefore, one can do a space-time split tailored to this frame. For example, an initial observer can be characterized by the tangent vector $\gamma_0$ to its world line. The vector $\gamma_0$ represents the observer's time axis, and any spacetime point, $x$, can be split into time and space components:
\begin{equation}
    x \gamma_0 ~=~ x\cdot\gamma_0 ~+~ x\wedge\gamma_0 ~=~ t ~+~ \bm{x}\,.
\end{equation}
Although $\bm{x} = x\wedge v$ is a space bivector, it can be treated as a vector in the $\gamma_0$-frame and is usually called the relative vector. 

The three-dimensional space in the $\gamma_0$ frame can be expanded by a standard frame of relative vectors:
\begin{equation}
    \bm{\sigma}_i = \gamma_i\gamma_0\,, ~\quad~i=1,2,3\,. 
\end{equation}
The $\{\sig{i}\}$ satisfy
\begin{equation}
    \sig{i}\cdot\sig{j} ~=~ \delta_{ij}\,.
\end{equation}
Furthermore, the volume element of the relative space is equal to the pseudoscalar in the spacetime:
\begin{equation}
    \sig{1}\sig{2}\sig{3} = \gamma_0 \gamma_1\gamma_2\gamma_3 = I\,.
\end{equation}
However, note that $I$ commutes with $\sig{i}$ and therefore can be considered as equivalent to $i=\sqrt{-1}$ in the relative space. 

The algebra generated by the $\{\bm{\sigma}_i\}$ is the even subalgebra of the STA $\mathcal{G}(3)$ and consists of 8 linearly independent elements:
\begin{equation}
    1\,, ~\quad~ \sig{i}\,, ~\quad~ I\sig{i}\,, ~\quad~ I\,.
\end{equation}
Furthermore, the reverse operation in this subalgebra works differently from the one in the STA and will be denoted by $\dagger$:
\begin{equation}
    \sig{i}^\dagger = \sig{i}\,, ~\quad~ (\sig{i}\sig{j})^\dagger = \sig{j}\sig{i}\,, ~\quad~ I^\dagger = \sig{3}\sig{2}\sig{1} = - I\,.
\end{equation}
Note that this operation corresponds to the Hermitian conjugate in standard quantum mechanics. The relation between this operation and the spacetime reversion is:
\begin{equation}
    M^\dagger = \gamma_0 \tilde{M} \gamma_0\,.
\end{equation}

%%%%%%%%%%%%%%%%%%%%%%%%%%%%%%%%%%%%%%%%%%%%%%%%%%%%%%%%%%%%%%%%%%%%%%%

\section{\label{sec:Bloch_non_Hermitian} BLOCH SPHERE EXTENSION TO THE LIGHT CONE}

It is well known that the pure state space of a two-level quantum system (qubit) can be represented geometrically by a unit 2-sphere called the Bloch sphere. To see this clearly, it is easier to check the density operator of a generic qubit:
\begin{equation}
    \rho ~=~ \frac{\bm{I} ~+~ n_x\,\sig{x} ~+~ n_y\,\sig{y} ~+~ n_z\,\sig{z}}{2}\,,
\end{equation}
where $\bm{I}$ is the identity and $\sig{x}$, $\sig{y}$ and $\sig{z}$ are Pauli matrices. The scalars $n_x$, $n_y$ and $n_z$ are the components of the Bloch vector $\vec{n}$, and they satisfy $n_x^2+n_y^2+n_z^2 = 1$. Therefore, the state can be represented by a unit vector called the Bloch vector, and the pure state space is the Bloch sphere.

Since the algebra of the Pauli matrices is isomorphic to the even subalgebra $\mathcal{G}(3)$, it is not surprising that there is a corresponding expression in the STA formulation. Once we choose an observer (i.e., the $\gamma_0$-frame), the density operator can be expressed with the relative vectors of this frame:
\begin{equation}
    \rho ~=~ \frac{1 ~+~ \hat{\bm{n}}}{2}\,,
\end{equation}
where $\hat{\bm{n}}$ is a unit vector in the relative space and its direction can be considered intuitively as the direction of the spin. Furthermore, the trace operation of the matrix formulation can be replaced by an extracting operation in the STA:
\begin{equation}
    \Tr({\rho}) ~\rightarrow~ \mathcal{N}\langle \rho \rangle\,,
\end{equation}
where $\mathcal{N}$ is the number of states that have been traced (i.e., $\mathcal{N}=2$ for a qubit) and the operation $\langle \_ \rangle$ extracts the scalar (i.e., grade-0) part. 

In the matrix formulation, the time-evolution of the density operator is:
\begin{equation}
    \rho(t) ~=~  U(t)\,\rho(0)\,U^\dagger(t)\,,
\end{equation}
where the time evolution operator $U(t)$ for a system with a time-independent Hamiltonian $H$ is:
\begin{equation}
    U(t) = e^{-i H t/\hbar}\,.
\end{equation}
In the STA formulation, the Hamiltonian $H$ is in $\mathcal{G}$(3) and contains theoretically grade-0, 2, and 4 elements. The grade-0 part does not affect the time evolution. On the other hand, the grade-4 part is an imaginary number in standard quantum mechanics, and it will induce an exponentially increasing or decreasing scalar factor on $\rho$ and render the dynamics non-unitary. Naturally, if we require the Hamiltonian to be Hermitian, the grade-4 part is not allowed. Nevertheless, since the effects of both the grade-0 and grade-4 parts are trivial (i.e., only introducing a scalar factor), we will focus mostly on the bivector part in this work. 

Furthermore, the Hermiticity of the Hamiltonian imposes further constraints on the allowed bivector. Note that the Hermitian conjugate corresponds to the reversion in the relative space in the STA; therefore, the Hermiticity of $H$ requires:
\begin{equation}
    H = H^\dagger = \gamma_0\, \tilde{H} \gamma_0\,.
\end{equation}
This is true when $H$ contains only the \emph{real} relative vector (i.e., time-like bivector in the STA) besides the scalar. Since $H$ is effectively a bivector, the time evolution operator can be recognized as a rotor:
\begin{equation}
    R = \exp(-B/2)\,,
\end{equation}
where $B$ is a bivector in the STA. Therefore, the time evolution of the density operator in a Hermitian system corresponds to a rotation in the relative space in the STA:
\begin{equation}
    R\, \rho\, R^\dagger = R\, \rho\, \tilde{R} ~=~ e^{-I \bm{b}/2}\,\rho\, e^{I \bm{b}/2} = \frac{1}{2}\,(1+e^{-I \bm{b}/2}\hat{\bm{n}}e^{I \bm{b}/2}) = \frac{1}{2}(1+\hat{\bm{n}}')\,,
\end{equation}
where $\bm{b}$ is a relative vector and the rotation is around the axis along the direction of $\bm{b}$.

So far, standard quantum mechanics requires only the even subalgebra of the STA. However, we will argue that to generalize to the system with the non-Hermitian Hamiltonian, it is more natural to lift the formulation to the full spacetime version. At first, note that we can write the density operator in the following way:
\begin{equation}
    \rho ~=~ \frac{1 ~+~ \hat{\bm{n}}}{2} = \frac{1}{2}\,n\,\gamma_0\,,
\end{equation}
where $n$ is a null vector (i.e., $n^2=0$) and $\hat{\bm{n}}=n\wedge\gamma_0$. The trace condition requires:
\begin{equation}
    \Tr(\rho) = 1 ~\rightarrow~ 2\langle \rho \rangle = \,n\cdot\gamma_0=1\,.
\end{equation}
Now the time evolution of the density operator is:
\begin{equation}
    \rho(t) = R(t)\, \rho\, R^\dagger(t) = \frac{1}{2} R(t)\,n\,\tilde{R}(t)\,\gamma_0 = \frac{1}{2}n'\gamma_0\,,
    \label{eq:rho_t}
\end{equation}
where the transformation
\begin{equation}
    n ~\rightarrow~ n' = R(t)\,n\,\tilde{R}(t) = e^{-B(t)/2}\,n\,e^{B(t)/2}\,,
\end{equation}
is a proper orthochronous Lorentz transformation on the null vector $n$ for a generic bivector $B$. This show that the state space of a non-Hermitian qubit is the future light cone. The general Lorentz transformation can be a boost, a rotation, or a mixture of both, depending on $B$. We will proceed with the analysis of the structure of the bivector $B$ in the section \ref{sec:Types_Lorentz}. 

The non-Hermitian dynamics is non-unitary, and therefore the probability is not necessarily conserved. This can be checked by verifying the trace of the density operator:
\begin{equation}
    \Tr({\rho'}) ~\rightarrow~ 2\langle\rho'\rangle = \langle e^{-B/2}\,n\,e^{B/2}\,\gamma_0\,\rangle =\langle \,n\,e^{B/2}\,\gamma_0\,e^{-B/2}\rangle\,.
    \label{eq:Tr_rho_general}
\end{equation}
The result can be seen as the inner product of the transformed null vector with $\gamma_0$ or the inverse-transformed $\gamma_0$ with the original null vector. If the Hamiltonian is Hermitian, the transformation is just a rotation in the relative space of the $\gamma_0$-frame. Naturally, the inverse rotation is also in the relative space and will not change $\gamma_0$. Therefore, the probability is conserved. However, for a more general spatial rotation, the total probability will oscillate with the period determined by the rotation speed and for a pure boost, the probability will either monotonically increase or decrease.

%%%%%%%%%%%%%%%%%%%%%%%%%%%%%%%%%%%%%%%%%%%%%%%%%%%%%%%%%%%%%%%%%%%%%%%%%%%%%%%%%%%%%%%%%%%%%%%%%%%%%%%%%%%%%%%%%%%%%%%%%%%%%%%%%%%%%

\section{\label{sec:Types_Lorentz}Geometric Classification and Types of Lorentz transformation}

In the previous section, we have shown that the non-Hermitian dynamics is related to a 'proper orthochronous’ Lorentz transformation of a null vector (up to some rescaling). The nature of this transformation is depending on the value of $B(t)^2$ which can be used to perform an Lorentz invarant decompoisition of $B(t)$ as mentioned in Chapter 5 of \cite{doran2003geometric}. In the following, we will mostly follow the work in \cite{doran2003geometric} and also assume $\hbar=1$ and set $t=1$ in this section for simplification, since they only rescale $B$ and will not change the conditions.

%To further understand the relation between the structure of $H$ and the corresponding Lorentz transformation, we need to analyze the properties of the bivector $B(t)$, and we will mostly follow the work in Chapter 5 of \cite{doran2003geometric}. However, before we proceed, we will assume $\hbar=1$ and set $t=1$ in this section to simplify the relation between $H$ and $B$, since they only rescale $B$ and will not change the conditions.

At first, note that the square of a bivector will contain only the grade-0 and grade-4 parts, and therefore:
\begin{equation}
    B^2 = \beta\,e^{I\,\phi}\,,
    \label{eq:B2_complex_parametrized}
\end{equation}
We can determine the nature of the transformation based on the result of $B^2$:
\begin{eqnarray*}
    &B^2 \in \mathbb{R}^+\, &~\rightarrow~ \textbf{Pure boosts}\,, \\
    &B^2 \in \mathbb{R}^-\, &~\rightarrow~ \textbf{Spatial rotations}\,, \\
    &B^2 = 0 &~\rightarrow~ \textbf{Null rotations}\,, \\
    &B^2 \in \mathbb{C}\, &~\rightarrow~ \textbf{General mixture of boost and rotation}\,.
\end{eqnarray*}
Note that $B^2=0$ is a special case corresponding to an exceptional point, which will be explained later. In the following decomposition procedure, we will assume $B^2\neq 0$. Based on the result of $B^2$, the bivector $B$ can be decomposed in a Lorentz-invariant way into the rotation and boost parts. At first, we can define:
\begin{equation}
    \hat{B} = \beta^{-1/2}\,e^{-I\phi/2}\,B\,,
\end{equation}
where $\hat{B}^2=1$. Then, $B$ can be decomposed into two bivector blades:
\begin{equation}
    B = \beta^{1/2}\,e^{I\phi/2}\,\hat{B} = \mu\,\hat{B} + \nu\,I\hat{B}\,.
    \label{eq:B_to_hatB}
\end{equation}
Since the two blades commute, the rotor $R$ decomposes into:
\begin{equation}
    R = e^{-\mu\hat{B}/2}\,e^{-\nu I\hat{B}/2} =  e^{-\nu I\hat{B}/2}\,e^{-\mu \hat{B}/2}\,,
\end{equation}
where $\hat{B}$ generate the boost and $I\hat{B}$ generate the rotation. 

Furthermore, for every time-like bivector $\hat{B}$, there is a pair of null vectors $n_{\pm}$ satisfying
\begin{equation}
    \hat{B}\cdot n_{\pm} = \pm n_{\pm}\,.
    \label{eq:B_npm}
\end{equation}
These two null vectors are in fact related to the two eigenstates of the qubit system and can be chosen such that
\begin{equation}
    n_+\wedge n_- = 2\hat{B}\,.
\end{equation}
The null vector $n_{\pm}$ anticommute with $\hat{B}$ and therefore commute with $I\hat{B}$.

If we ignore the grade-0 and grade-4 parts, the most general Hamiltonian can be written as:
\begin{equation}
    H = \bm{a} + I \bm{b}\,,
\end{equation}
where $\bm{a}$ and $\bm{b}$ are two generic relative vectors. Furthermore, assuming $\hbar=1$ and $t=1$, the relation between the Hamiltonian $H$ and $B$ is simply:
\begin{equation}
    B = 2 IH\,.
\end{equation}
Therefore, the square of $B$ is:
\begin{equation}
    B^2 = - 4 H^2 = -4 (\bm{a} + I \bm{b})^2 = -4(\bm{a}^2 - \bm{b}^2) - 8 I(\bm{a}\cdot\bm{b})\,.
    \label{eq:B2_cond}
\end{equation}
Therefore the nature of the corresponding Lorentz transform will depend on $\bm{a}\cdot \bm{b}$ and the relative length of $\bm{a}$ and $\bm{b}$. In the following, we discuss four different types of transformation and their conditions.

%Since the bivector decomposes into two commuting blades, the most general form of the bivector can be written as:
%\begin{equation}
%    B = \bm{a} + I\bm{b}\,,
%\end{equation}
%where $\bm{a}$ and $\bm{b}$ are some generic relative vectors of the $\gamma_0$-frame. The square of $B$ becomes
%\begin{equation}
%    B^2 = (\bm{a} + I\bm{b})^2 = \bm{a}^2-\bm{b}^2 + 2 I(\bm{a}\cdot\bm{b})\,.
%\end{equation}
%This implies that if $\bm{a}$ and $\bm{b}$ are orthogonal, the transformation will be either pure boost or pure rotation depending on the relative lengths of $\bm{a}$ and $\bm{b}$. However, the transformation will be a mixture if the two relative vectors are not orthogonal. In the following, we discuss several different types of transformation based on the values of $B^2$.

\subsection{Spatial rotation}

\begin{figure}
    \centering
    \includegraphics[width=0.5\linewidth]{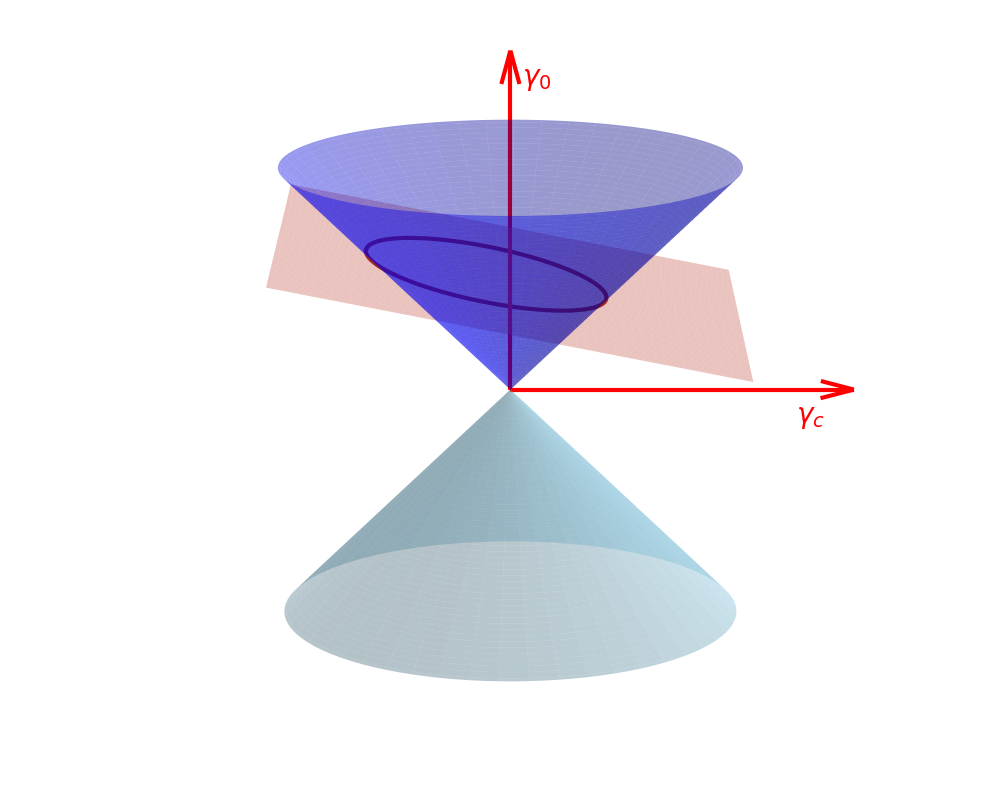}
    \caption{The light cone and tilted rotation generated by a non-Hermitian Hamiltonian satisfied the conditions in (\ref{eq:p_rotaion_cond}). The rotation is on the tilted plane because $\gamma_0'$ is tilted toward $\gamma_c$ direction. The cross-section of the tilted plane with the light cone is the trajectory of the null vector under the transformation.}
    \label{fig:tilt_rotation}
\end{figure}

For the transformation to be a spatial rotation, the square of $B$ must be a real negative value. From the equation (\ref{eq:B2_cond}), we see that the conditions are:
\begin{equation}
    \bm{a}^2 > \bm{b}^2\,,~\quad~\bm{a}\cdot\bm{b}=0\,.
    \label{eq:p_rotaion_cond}
\end{equation}
Note that if the Hamiltonian $H$ is Hermitian, $\bm{b}=0$ and therefore, the above conditions are satisfied trivially. Since the orthogonal condition removes one degree of freedom, the corresponding $H$ in this case is parametrized by five parameters, subject to an additional constraint: $\bm{a}^2>\bm{b}^2$. Including the real scalar in the Hamiltonian, this is exactly the six-parameter parametrization of the general $\mathcal{PT}$-symmetric Hamiltonian for a two-level system in \cite{wang2010symmetry}. Assuming the conditions in (\ref{eq:p_rotaion_cond}) are satisified, the bivector $B$ can be written:
\begin{equation}
    B = 2 \sqrt{\bm{a}^2-\bm{b}^2}\,I\hat{B}\,,
\end{equation}
where $\hat{B}^2=1$ and
\begin{equation}
    \hat{B} = \frac{1}{\sqrt{\bm{a}^2-\bm{b}^2}}\,(\bm{a} + I\bm{b})\,.
\end{equation}
Furthermore, we can express $\hat{B}$ in the STA bivector form:
\begin{equation}
    \hat{B} = \gamma_a\,\gamma_0'\,, 
    \label{eq:B_hat_pure_rotation_STA}
\end{equation}
where $\gamma_0'$ is a new unit time-like vector:
\begin{equation}
    \gamma_0'= \left(\sqrt{\frac{\bm{a^2}}{\bm{a}^2-\bm{b}^2}}\,\gamma_0 ~+~ \sqrt{\frac{\bm{b^2}}{\bm{a}^2-\bm{b}^2}}\,\gamma_c\right)\,.
\end{equation}
The $\gamma_{a,b,c}$ will denote the three orthonormal space-like vector such that:
\begin{equation}
    \gamma_{a} = \hat{\bm{a}}\gamma_0\,, ~\quad~ \gamma_{b} = \hat{\bm{b}}\gamma_0\,, ~\quad~ \gamma_{b} = \hat{\bm{c}}\gamma_0\,,
\end{equation}
where $\hat{\bm{a}}$ and $\hat{\bm{b}}$ are the unit relative vectors point along $\bm{a}$ and $\bm{b}$ respectively and $\hat{\bm{a}}$ and $\hat{\bm{b}}, \hat{\bm{c}}$ form a right-handed frame: $I=\hat{\bm{a}}\hat{\bm{b}}\hat{\bm{c}}=\gamma_0\gamma_a\gamma_b\gamma_c$. Note that for a Hermitian Hamiltonian, $\gamma_0'=\gamma_0$, and it is precisely the non-Hermiticity of $H$ that kind of boosts the $\gamma_0$ in the direction of $\gamma_c$. We can also check that the probability is not conserved. From the equation (\ref{eq:Tr_rho_general}), the total probability is:
\begin{equation}
    \langle e^{-\sqrt{\bm{a}^2-\bm{b}^2}I\hat{B}} n e^{\sqrt{\bm{a}^2-\bm{b}^2}I\hat{B}}\gamma_0 \rangle ~=~\langle n e^{\sqrt{\bm{a}^2-\bm{b}^2}I\hat{B}}\gamma_0 e^{-\sqrt{\bm{a}^2-\bm{b}^2}I\hat{B}}\rangle
\end{equation}
Note that the transformation generated by $I\hat{B}$ is kind of a tilted rotation that is not aligned with the relative space of the $\gamma_0$-frame as shown in the figure \ref{fig:tilt_rotation}. The trajectory of a null vector is an ellipse in this case. Furthermore, from the figure, we can see that the probability will not be fixed but oscillate with a fixed period.

The eigenstates of this system can be obtained from the two invariant null vectors:
\begin{equation}
    n_{\pm} = \sqrt{\frac{\bm{a}^2-\bm{b}^2}{\bm{a}^2}}\left(\gamma_0' \pm\gamma_a\right)\,.
    \label{eq:eigen_null_rotation}
\end{equation}
The scalar factor is just for preserving $\Tr({\rho})=1$. The corresponding density operators are:
\begin{equation}
    \rho_{\pm} = \frac{1}{2}n_{\pm}\gamma_0\,.
\end{equation}
Their corresponding eigenvalues can be obtained from:
\begin{equation}
    H\rho_{\pm}= \frac{1}{2}\sqrt{\bm{a}^2-\bm{b}^2}\,\hat{B}n_{\pm}\gamma_0 = \pm\sqrt{\bm{a}^2-\bm{b}^2}\rho_{\pm}\,,
\end{equation}
where we use the fact that $n_{\pm}$ satisfy (\ref{eq:B_npm}). On the other hand, we can also calculate the expectation values of the Hamiltonian for the two eigenstates:
\begin{equation}
    \Tr(\rho_{\pm}H) ~\rightarrow~ 2\langle\rho_{\pm}H\rangle = \langle n_{\pm}\gamma_0 H \rangle = \sqrt{\bm{a}^2-\bm{b}^2}\,\langle n_{\pm}\gamma_0 \hat{B}\rangle = \pm\sqrt{\bm{a}^2-\bm{b}^2}\,.
\end{equation}
Therefore, in the case of a spatial rotation, the eigenvalues of $H$ are real even though the Hamiltonian is non-Hermitian. 

Note that from the form of $\hat{B}$ in (\ref{eq:B_hat_pure_rotation_STA}), it is clear that if we can boost $\gamma_0'$ back to $\gamma_0$, we can restore the Hermiticity of the Hamiltonian $H$. The rotor for this boost is not hard to find:
\begin{equation}
    R_{\hat{\bm{c}}} = e^{-\alpha\hat{\bm{c}}/2} = \cosh(\alpha/2) - \sinh(\alpha/2)\hat{\bm{c}}\,, 
    \label{eq:R_boost_back}
\end{equation}
where $\alpha$ is a boost parameter such that:
\begin{equation}
    \cosh(\alpha) = \sqrt{\frac{\bm{a^2}}{\bm{a}^2-\bm{b}^2}}\,, ~\quad~\sinh(\alpha)=\sqrt{\frac{\bm{b^2}}{\bm{a}^2-\bm{b}^2}}\,.
\end{equation}
Under this boost transformation, the Hamiltonian $H$ becomes:
\begin{equation}
    H ~\rightarrow~H'= R_{\hat{\bm{c}}} H \tilde{R}_{\hat{\bm{c}}} = R_{\hat{\bm{c}}}(\bm{a}+I\bm{b})\tilde{R}_{\hat{\bm{c}}} = \sqrt{\bm{a}^2-\bm{b}^2}\,\gamma_a\,R_{\hat{\bm{c}}}\gamma_0'\tilde{R}_{\hat{\bm{c}}}=\sqrt{\bm{a}^2-\bm{b}^2}\,\hat{\bm{a}}\,.
    \label{eq:H_boost}
\end{equation}
Now the new Hamiltonian $H'$ is Hermitian. Since the boost transforms $\gamma_0'$ in $\hat{B}$ to $\gamma_0$, the two invariant null vectors $n_{\pm}$ must also change accordingly. Therefore, the eigenstates will be different, but the spectrum remains the same.

We can also further compare the formulation in the STA with the $\mathcal{PT}$-symmetric matrix formulation in \cite{wang2010symmetry}. We will list the corresponding elements in the STA for several operators defined in \cite{wang2010symmetry} and show that the result of two-level systems matches the analysis in the STA formulation.

The parity operation $\mathcal{P}$ corresponds to $\hat{\bm{a}}$. The time reversal operator $\mathcal{T}$ is the Hermitian conjugate and therefore corresponds to the reversion in the relative space. Naturally, the Hamiltonian $H$ is $\mathcal{PT}$-symmetric:
\begin{equation}
    \hat{\bm{a}}H^\dagger\hat{\bm{a}}=\hat{\bm{a}}(\bm{a}-I\bm{b})\hat{\bm{a}} = \bm{a} + I\bm{b} =H\,,
\end{equation}
as long as $\bm{a}$ and $\bm{b}$ are orthogonal. The $\mathcal{C}$ operator corresponds to $\hat{B}$ and the weight function $W$ corresponds to:
\begin{equation}
    W=\mathcal{PC} ~\rightarrow~ \hat{\bm{a}}\hat{B} = \left(\sqrt{\frac{\bm{a}^2}{\bm{a}^2-\bm{b}^2}}-\sqrt{\frac{\bm{b}^2}{\bm{a}^2-\bm{b}^2}}\hat{\bm{c}} \right)\,.
\end{equation}
This is in fact a rotor for boosting, and it transforms $H$ to $H^\dagger$:
\begin{equation}
    (\hat{\bm{a}}\hat{B})\,H\,(\hat{B}\hat{\bm{a}}) = H^\dagger\,.
\end{equation}
The square root $\eta$ (i.e., $W=\eta^2$) corresponds to the rotor $R_{\hat{\bm{c}}}$ in (\ref{eq:R_boost_back}) that boosts $\gamma_0'$ back to $\gamma_0$. Naturally, the transformation in \cite{wang2010symmetry},
\begin{equation}
    H ~\rightarrow~ h=\eta\,H\eta^{-1}\,,
\end{equation}
is precisely the boost defined in (\ref{eq:H_boost}).

Furthermore, we can also check the $\mathcal{CPT}$ inner product defined in \cite{wang2010symmetry}:
\begin{equation}
    (\psi,\phi)_{\mathcal{CPT}}\equiv\langle\psi|\mathcal{PC}|\phi\rangle\,.
\end{equation}
From the definition, the $\mathcal{CPT}$-norms of the eigenstates $\rho_{\pm}$ in the STA is:
\begin{equation}
    \Tr(\mathcal{PC}\rho_{\pm})  ~\rightarrow~ \langle R_{\hat{\bm{c}}}^2\, n_{\pm}\gamma_0\rangle = \langle R_{\hat{\bm{c}}}\, n_{\pm}\tilde{R}_{\hat{\bm{c}}}\gamma_0\rangle\,. 
\end{equation}
The boost transform $\gamma_0'$ in $n_{\pm}$ back to $\gamma_0$ and therefore, if we define $n_{\pm}=(\gamma_0'\pm\gamma_a)$, the $\mathcal{CPT}$-norms of the eigenstates are normalized to one. Note that these two eigenstates are not orthogonal in the original inner product:
\begin{equation}
    \Tr(\rho_+\rho_-) ~\rightarrow~ \langle n_+\gamma_0\, n_-\gamma_0\rangle = \langle (\gamma_0'+\gamma_a)(\gamma_0''+\gamma_a)\rangle \neq 0\,,
\end{equation}
where $\gamma_0''$ is
\begin{equation}
    \gamma_0'' = \gamma_0\gamma_0'\gamma_0 = \left(\sqrt{\frac{\bm{a^2}}{\bm{a}^2-\bm{b}^2}}\,\gamma_0 ~-~ \sqrt{\frac{\bm{b^2}}{\bm{a}^2-\bm{b}^2}}\,\gamma_c\right)\,.
\end{equation}
Therefore, the two eigenstates will not be orthogonal in the original inner product unless the Hamiltonian is Hermitian. However, the two states are indeed orthogonal in the $\mathcal{CPT}$ inner product:
\begin{equation}
    \Tr(\mathcal{PC}\rho_+\rho_-) ~\rightarrow~ \langle R_{\hat{\bm{c}}}^2\, n_+\gamma_0\,n_-\gamma_0 \rangle =  \langle R_{\hat{\bm{c}}}^2 (\gamma_0'+\gamma_a)(\gamma_0''+\gamma_a)\rangle =\langle R_{\hat{\bm{c}}}\gamma_0' \tilde{R}_{\hat{\bm{c}}}\gamma_0''\rangle - \langle R_{\hat{\bm{c}}}^2\rangle = 0\,.
\end{equation}

\subsection{Pure boost}

\begin{figure}
    \centering
    \includegraphics[width=0.5\linewidth]{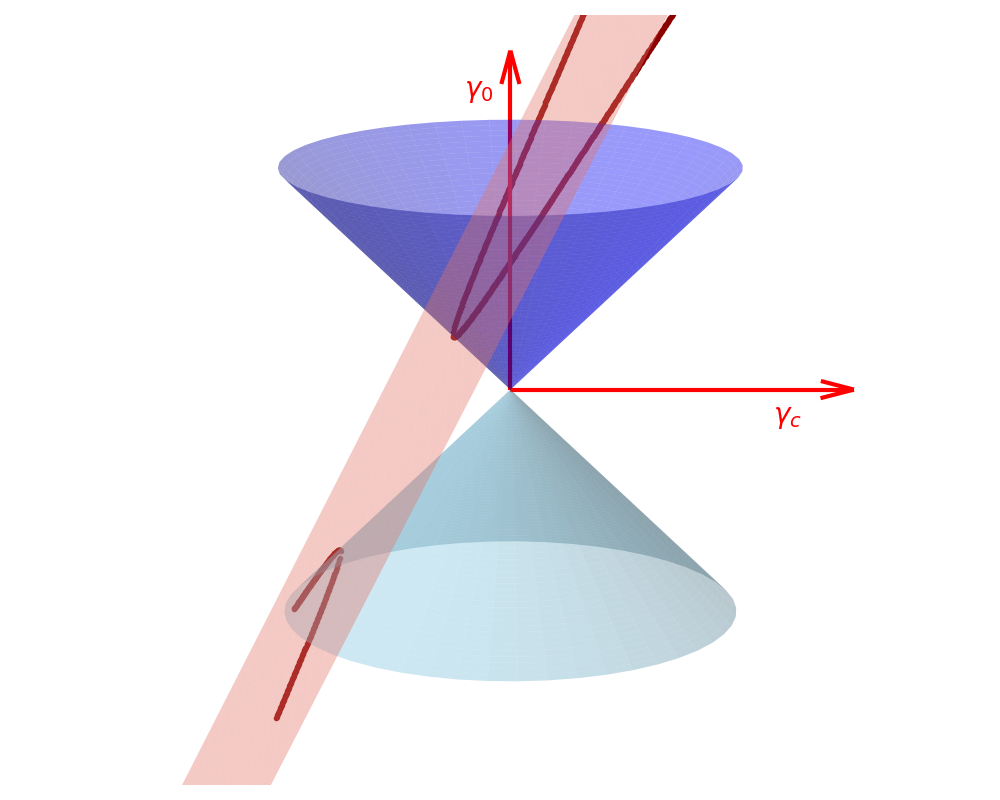}
    \caption{The light cone and tilted rotation generated by a non-Hermitian Hamiltonian satisfied the conditions in (\ref{eq:p_boost_cond}). The boost is on the tilted plane because $\gamma_0'''$ is tilted toward $\gamma_c$ direction. The cross-section of the tilted plane with the future light cone is the trajectory of the null vector under the transformation.}
    \label{fig:tilt_boost}
\end{figure}

For the transformation to be a pure boost, the square of $B$ must be real and positive. From the equation (\ref{eq:B2_cond}), we see that the conditions are:
\begin{equation}
    \bm{a}^2 < \bm{b}^2\,,~\quad~\bm{a}\cdot\bm{b}=0\,.
    \label{eq:p_boost_cond}
\end{equation}
The bivector $B$ can be written:
\begin{equation}
    B = 2 \sqrt{\bm{b}^2-\bm{a}^2}\,\hat{B}\,,
\end{equation}
where $\hat{B}^2=1$ and
\begin{equation}
    \hat{B} = \frac{1}{\sqrt{\bm{b}^2-\bm{a}^2}}\,IH = \frac{1}{\sqrt{\bm{b}^2-\bm{a}^2}}\, (-\bm{b} + I\bm{a})\,.
\end{equation}
Furthermore, we can express $\hat{B}$ in the STA bivector form:
\begin{equation}
    \hat{B} = -\gamma_b\,\gamma_0'''\,, 
    \label{eq:B_hat_pure_rotation_STA}
\end{equation}
where $\gamma_0'''$ is a unit time-like vector:
\begin{equation}
    \gamma_0''' = \left(\sqrt{\frac{\bm{b^2}}{\bm{b}^2-\bm{a}^2}}\,\gamma_0 ~+~ \sqrt{\frac{\bm{a^2}}{\bm{b}^2-\bm{a}^2}}\,\gamma_c\right)\,.
\end{equation}
Therefore, the time evolution of this case will be a kind of "tilted" boost as shown in the figure \ref{fig:tilt_boost}. The trajectory of a null vector is an hyperbolic in this case. Furthermore, from the figure, we can see that the probability will not be conserved but grow or decay exponentially.

The two null eigenvectors are:
\begin{equation}
    n_{\pm} = \sqrt{\frac{\bm{b}^2-\bm{a}^2}{\bm{b}^2}}(\gamma_0''' \mp \gamma_b)\,.
    \label{eq:eigen_null_boost}
\end{equation}
The eigenstates can be constructed from them and the eigenvalues can be derived from:
\begin{equation}
    H\rho_{\pm} = -I\sqrt{\bm{b}^2-\bm{a}^2}\hat{B}\,\left(\frac{1}{2}n_{\pm}\gamma_0\right)=\mp I \sqrt{\bm{b}^2-\bm{a}^2}\rho_{\pm}\,.
\end{equation}
Therefore, the eigenvalues of the Hamiltonian are pure imaginary numbers:
\begin{equation}
    E_{\pm} = \mp I\sqrt{\bm{b}^2-\bm{a}^2}\,.
\end{equation}
This corresponds to the anti-$\mathcal{PT}$ symmetric case in the literature.

Furthermore, we can also try to find the boost that transforms $\gamma_0'''$ back to $\gamma_0$:
\begin{equation}
    R'_{\hat{\bm{c}}}(\gamma_0''')\tilde{R}'_{\hat{\bm{c}}} = \gamma_0\,,
\end{equation}
where $(R'_{\hat{\bm{c}}})^2$ is
\begin{equation}
    (R'_{\hat{\bm{c}}})^2 = \left(\sqrt{\frac{\bm{b}^2}{\bm{b}^2-\bm{a}^2}} - \sqrt{\frac{\bm{a}^2}{\bm{b}^2-\bm{a}^2}} \hat{\bm{c}}\right)   = -\hat{\bm{b}}\hat{B}\,.
\end{equation}
This boost will also transform $H$ to an anti-Hermitian Hamiltonian:
\begin{equation}
    R'_{\hat{\bm{c}}} H\tilde{R}'_{\hat{\bm{c}}} = -I\sqrt{\bm{b}^2-\bm{a}^2}\gamma_b(-\gamma_0) = \sqrt{\bm{b}^2-\bm{a}^2}\,I\hat{\bm{b}}\,. 
\end{equation}

\subsection{Null rotation}

When $\bm{a}^2=\bm{b}^2$ and $\bm{a}\cdot\bm{b}=0$, $B^2=0$. The corresponding transformation is referred to as a null rotation. A null bivector can be written as a geometric product of a null vector and an orthogonal space-like vector. The Hamiltonian $H$ in this case is null and can be written as:
\begin{equation}
    H =\bm{a} + I\bm{b} = \sqrt{\bm{a}^2}\gamma_a(\gamma_0+\gamma_c) = a\,n_{c}\,, 
\end{equation}
where $a=\bm{a}\gamma_0$ and $n_c$ is a null vector. Note that in this situation, there is only one null eigenvector with zero eigenvalue:
\begin{equation}
    H n_c = a\,n_c^2=0\,.
\end{equation}

As $\bm{a}^2 - \bm{b}^2 \to 0$ (while preserving $\bm{a}\cdot\bm{b}=0$), the two null eigenvectors associated with both spatial rotations (\ref{eq:eigen_null_rotation}) and pure boosts (\ref{eq:eigen_null_boost}) coalesce into a single null direction $n_c$. This state coalescence directly signals the approach to an exceptional point, establishing that null rotations in STA geometrically represent exceptional points in non-Hermitian quantum mechanics. 

A null vector will transform under the null rotation:
\begin{equation}
    n \rightarrow n' = e^{-B/2}\,n\,e^{B/2} =(1-\frac{B}{2})\,n\,(1+\frac{B}{2}) = n-I(H n+nH) - HnH\,.
\end{equation}
In this regime, the state traces a parabolic trajectory on the light cone, causing the overall probability to grow or decay polynomially over time.

\subsection{Mixture of boost and rotation}

\subsubsection{A simplified example}

The situation will become more complicated if $\bm{a}\cdot\bm{b}\neq0$. In this case, the transform is a mixture of boost and rotation. From (\ref{eq:B2_complex_parametrized}) and (\ref{eq:B2_cond}), the parameters $\beta$ and $\phi$ can be related to $\bm{a}$ and $\bm{b}$:
\begin{equation}
    \beta\cos(\phi) = -4(\bm{a}^2-\bm{b}^2)\,,~\quad~ \beta\sin(\phi)= -8(\bm{a}\cdot\bm{b})\,.
\end{equation} 
One can proceed as usual, but the computation will become quite complicated. To illustrate the general features of this case without overcomplicating, we will use a simplified example. Particularly, consider a Hamiltonian H with $\bm{a}^2=\bm{b^2}$ but $\bm{a}\cdot\bm{b}>0$. Then, we have:
\begin{equation}
    \beta = 8(\bm{a}\cdot\bm{b})\,, ~\quad~\phi=3\pi/2\,.
\end{equation}
The unit time-like bivector $\hat{B}$ is:
\begin{equation}
    \hat{B} = \frac{1}{2\sqrt{2(\bm{a}\cdot \bm{b)}}}\,e^{-I\frac{3}{4}\pi}\,B=\frac{1}{\sqrt{2(\bm{a}\cdot \bm{b)}}}\,e^{-I\frac{3}{4}\pi}\,(I\bm{a}-\bm{b})\,.
\end{equation}
It can also be expressed in the bivector form of the STA:
\begin{equation}
    \hat{B} = \gamma_+\gamma_0^*\,,
\end{equation}
where $\gamma_0^*$ is another unit timelike vector:
\begin{equation}
    \gamma_0^* = \left(\frac{1}{2}\sqrt{\frac{(\bm{a} +\bm{b})^2}{\bm{a}\cdot \bm{b}}}\gamma_0+\frac{1}{2}\sqrt{\frac{(\bm{a}-\bm{b})^2}{\bm{a}\cdot \bm{b}}}\gamma_c \right)\,,
\end{equation}
and the unit space-like vector $\gamma_{\pm}$ point at the direction of $(\bm{a}\pm\bm{b})\gamma_0$. The handedness is defined as $I=\gamma_0\gamma_-\gamma_+\gamma_c$. The two eigen null vectors are:
\begin{equation}
    n_{\pm} ~\propto~ \gamma_0^* \pm\gamma_+\,.
\end{equation}

The Hamiltonian $H$ can be expressed with $\hat{B}$:
\begin{equation}
    H = -\frac{I}{2}B=-\frac{I}{2}\beta^{1/2}e^{I\phi/2}\hat{B}=\sqrt{\bm{a}\cdot\bm{b}}\,(1+I)\hat{B}\,.
\end{equation}
Since $\hat{B}$ act on $n_{\pm}$ will just yield $\pm 1$, the eigenvalues can be read from above easily:
\begin{equation}
    E_{\pm} = \pm\sqrt{\bm{a}\cdot\bm{b}}\,(1+I)\,,
\end{equation}
which are complex. In fact, from the relation between $B$ and $\hat{B}$ in (\ref{eq:B_to_hatB}), we can see that the eigenvalues are always complex if the transformation contains a mixture of boost and rotation.  

\subsubsection{Topological features of exceptional points(EP)}

It is well known that the parameter space near an exceptional point shows non-trivial topological structure. Specifically, an exceptional point acts as a branch-point singularity on a Riemann eigenvalue surface. The two eigenstates will interchange when one circles around a second-order EP once and only return to the original state after circling the second round. We can also investigate these aspects in the STA formulation. 

\begin{figure*}
\includegraphics[width=0.9\textwidth]{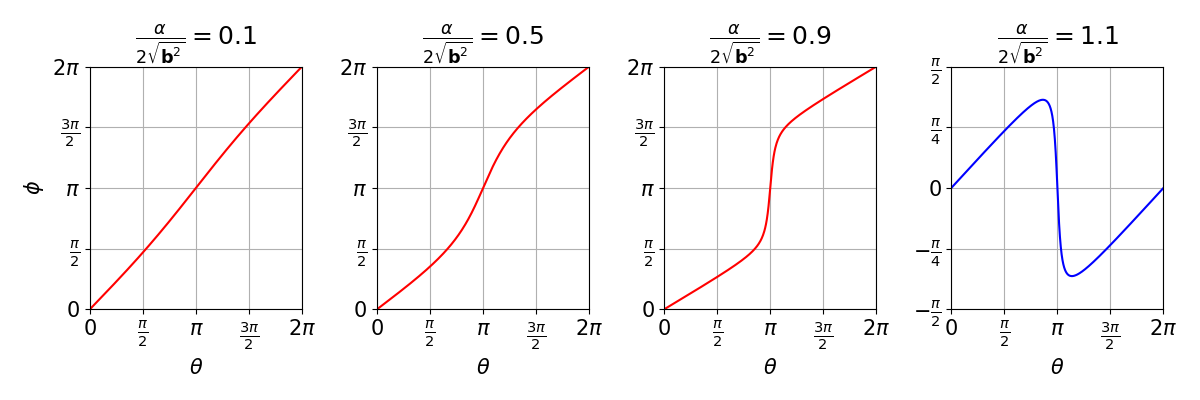}% Here is how to import EPS art
\caption{\label{fig:phi_theta}The figures show the relation between $\theta$ and $\phi$ for different values of $\frac{\alpha}{2\sqrt{\bm{b}^2}}$.}
\end{figure*}

Here we would like to demonstrate this topological feature in an example. Consider a Hamiltonian $H$ in which we fix $\bm{b}$ but leave $\bm{a}$ tunable. Since the conditions for exceptional points are $\bm{a}^2=\bm{b}^2$ and $\bm{a}\cdot\bm{b}=0$, they form a ring on the plane perpendicular to $\bm{b}$. We can parametrize $\bm{a}$ such that it circles an exceptional point on the ring, like:
\begin{equation}
    \bm{a}(\theta) = \sqrt{\bm{b}^2} \hat{\bm{a}'} + \alpha\cos(\theta)\hat{\bm{a}'} + \alpha\sin(\theta)\hat{\bm{b}}\,,
\end{equation}
where $\hat{\bm{b}}$ is the unit vector point at $\bm{b}$ and $\hat{\bm{a}'}\cdot \hat{\bm{b}}=0$. Then, we can compute $B^2$:
\begin{eqnarray}
    B^2 ~&=&~ -4(\bm{a}^2(\theta) - \bm{b}^2) - 8 I(\bm{a}(\theta)\cdot\bm{b})\,, \nonumber \\
        ~&=&~ -4(2 \sqrt{\bm{b}^2}\alpha\cos(\theta) + \alpha^2) - 8I\sqrt{\bm{b}^2}\alpha\sin(\theta).
\end{eqnarray}
Compare the above equation with (\ref{eq:B2_complex_parametrized}), we can obtain $\beta$:
\begin{equation}
    \beta ~=~ 4\,\alpha\sqrt{\alpha^2+4\alpha\sqrt{\bm{b}^2}\cos(\theta)+4\bm{b}^2}\,.
\end{equation}
And, the phase angle $\phi$ satisfies:
\begin{equation}
    \tan(\phi)=\frac{\sin(\theta)}{\cos(\theta) + \frac{\alpha}{2\sqrt{\bm{b}^2}}}\,.
\end{equation}
Clearly, if the radius of the circle $\alpha$ that we choose is much smaller than $\sqrt{\bm{b}^2}$, the phase angle $\phi\approx\theta$. Therefore, when the vector $\bm{a}$ circles the EP once, $\phi$ also completes a full circle from $0$ to $2\pi$. However, if we increase $\alpha$, the phase angle $\phi$ will deviate from $\theta$. Still, $\phi$ still complete a full circle if $\alpha<2\sqrt{\bm{b}^2}$. If the radius is long enough that $\alpha>2\sqrt{\bm{b}^2}$, $\phi$ will not complete a full circle as $\theta$ goes from $0$ to $2\pi$, as shown in Fig. \ref{fig:phi_theta}.
 
The eigenvalues of $H$ are:
\begin{equation}
    E_{\pm} = \mp\frac{I}{2}\beta^{1/2}\,e^{I\phi/2}\,.
\end{equation}
This form of the eigenvalues makes it clear why the eigenstates exchange when we go around the EP. If $\alpha<2\sqrt{\bm{b}^2}$, the phase $\phi$ completes a full cycle when $\bm{a}$ goes around the EP once. Consequently, the eigenvalues will not return to their original values but acquire a negative sign from $e^{I\pi}=-1$. Therefore, the two eigenvalues and eigenstates will exchange such that $E_{\pm} \rightarrow E_{\mp}$. On the other hand, if $\alpha > 2\sqrt{\bm{b}^2}$, the phase $\phi$ will not complete a full cycle; therefore, the eigenvalues will return to their original values. These two cases are topologically inequivalent, and they can not be deformed continuously to each other without crossing the EP point. Furthermore, we can see that this branch cut behavior near the EP is coming from the square root of $B^2$.

\section{\label{sec:discussion}Discussion and Future Directions}

%\textbf{Discussion}
%\begin{enumerate}
%    \item The physical meaning of this null vector hiding in the density operator.
%    \item Change of observers implies $n\gamma_0 \rightarrow nv$?

In this paper, we showed that the Bloch sphere can be generalized to systems with a non-Hermitian Hamiltonian. Just as the time evolution of a state can be understood as a rotation of the Bloch vector around a fixed axis on the Bloch sphere, it can be understood as a general Lorentz transformation of a null vector on the future light cone in a more general case. Depending on the Hamiltonian, the time evolution can be categorized into four different kinds corresponding to four types of Lorentz transformations. We also showed that the $\mathcal{PT}$-symmetric Hamiltonian corresponds to a spatial rotation, the anti-$\mathcal{PT}$ symmetric one corresponds to a pure boost, and the exceptional point corresponds to a null rotation. We also showed that several works in non-Hermitian quantum mechanics, including topological features of an exceptional point, can be studied more geometrically in this formulation. 

Although we have shown how the time evolution of a state connects to a Lorentz transformation of a null vector, it is still unclear whether there is any deeper meaning behind it. As we have shown previously, the density operator of a qubit can be written as:
\begin{equation}
    \frac{1}{2}n\gamma_0 ~=~ \frac{1}{2}\left(n\cdot\gamma_0 + n\wedge\gamma_0\right)\,,
\end{equation}
in the $\gamma_0$-frame. To preserve $\Tr(\rho)=1$, we divide the above equation by $n\cdot\gamma_0$ to obtain:
\begin{equation}
    \rho ~=~ \frac{1}{2}\left(1 + \frac{n\wedge\gamma_0}{n\cdot\gamma_0}\right) = \frac{1}{2}(1+\hat{\bm{n}})\,,
\end{equation}
where $\hat{\bm{n}}$ is a unit relative vector of $n$ in the $\gamma_0$-frame \cite{doran2003geometric}, . When we let time flow for some amount of time, the null vector $n$ can transform to $n' = R n \tilde{R}$ by a Lorentz transformation $R$. If we enforce the trace condition $\Tr(\rho)=1$, the unit relative vector becomes:
\begin{equation}
    \hat{\bm{n}} ~\rightarrow~ \hat{\bm{n}}'=\frac{n'\wedge\gamma_0}{n'\cdot\gamma_0} = R\frac{n\wedge v}{n\cdot v}\tilde{R}\,,
\end{equation}
where $v=\tilde{R}\gamma_0 R$. Therefore, the time-transformed unit relative vector $\hat{\bm{n}}'$ is equivalent to a non-transformed unit relative vector in the $v$-frame brought back to the $\gamma_0$-frame. This seems to imply the time evolution can be understood as a change of the observer. Furthermore, the Lorentz transformation can be viewed visually as a mapping of a point to another point on the celestial sphere \cite{doran2003geometric} (i.e, $\hat{\bm{n}}\rightarrow\hat{\bm{n}}'$). This seems to suggest that the time evolution of a Hamiltonian can be understood as a mapping in a two-dimensional space. Anyway, these are some open questions left to be explored, and we will leave them for a future investigation. 

Another important question is whether we can generalize this picture to multi-qubit cases. A possible way is to follow \cite{wang2018density} by using the multi-particle spacetime algebra (MSTA). However, how to treat the pseudoscalars from different particles consistently may become a subtle problem. Again, we will leave this problem to a future exploration.

\begin{acknowledgments}
The author would like to thank Prof. Jhih-Sheng Wu for making him aware of the development of non-Hermitian quantum mechanics.
\end{acknowledgments}

\bibliography{NH_Bloch}% Produces the bibliography via BibTeX.

\end{document}